# Extended Hamilton's principle

Jinkyu Kim

*Department of Mechanical and Aerospace Engineering*
*University at Buffalo, State University of New York*
*Buffalo, NY 14260 USA*

jk295@buffalo.edu

**Abstract**

Hamilton's principle is extended to have compatible initial conditions to the strong form. To use a number of computational and theoretical benefits for dynamical systems, the mixed variational formulation is preferred in the systems other than particle systems. With this formulation and the Rayleigh's dissipation function, we could have all the pertinent initial/boundary conditions for both conservative and non-conservative dynamical system. Based upon the extension framework of Hamilton's principle, the numerical method for representative lumped parameter models is also developed through applying Galerkin's method to time domain with the discussion of its numerical properties and simulation results.

## 1. Introduction

Dynamics of the system has a nature of integration in both space and time. Hamilton's Principle (Hamilton, 1834, 1835) may be a theoretical base for dynamical systems by its nature of integral form in time with Lagrangian density to account for continuous space. However, it has critical weakness, the end-point constraints, which imply that the positions of the dynamic system are known at the beginning and at the end of the time interval. Considering that the primary objective of initial value problems is to find the position in the future, how can we think that the position at the end time is known?
The main objective of the present work is to show how Hamilton's principle could be extended to circumvent such critical weakness. The paper is organized as follows.

In Section 2, Hamilton's principle is reviewed with pointing out its critical weakness and limit. Then, in Section 3, the previous works, which invoke the insight on end-point constraints and the mixed formulation of Hamilton's principle, are highlighted. Section 4 presents the extension framework of Hamilton's principle that can correctly account for initial value problems. As we shall see, the new framework presented there recovers the governing differential equations along with the specified initial and boundary conditions for elastodynamic continuua. With this new framework, the numerical method for the damped oscillator and the elastic viscoplastic dynamic system is developed in Section 5. There, the algorithm and the numerical properties for each model are also explained. Numerical simulation results for elastic viscoplastic dynamic system are shown in Section 6, and finally the work is summarized and some conclusions are drawn in Section 7.

## 2. Hamilton's principle

Hamilton's principle is an example of calculus of variations in mathematics (see Calkin, 1996; Fox, 1987; Gelfand and Fomin, 2000; Goldstein, 1980; Lanczos, 1986): The true trajectory of the system is found to make the functional, action, be stationary.

Consider the harmonic oscillator displayed in Fig. 1, consisting of a mass $m$ and linear spring having constant stiffness $k$.

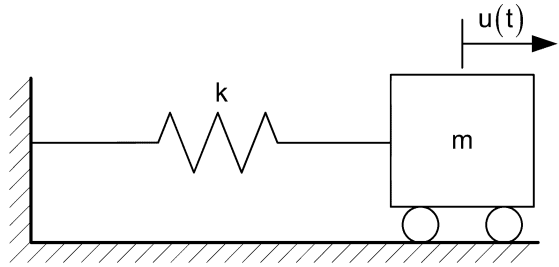

**Fig. 1. Harmonic oscillator**

The functional action $A$ for the fixed time interval from $t_0$ to $t_1$ is written

$$A(u,\dot{u};t) = \int_{t_0}^{t_1} L(u,\dot{u};\tau)\,d\tau \tag{1}$$

where the Lagrangian is

$$L(u,\dot{u};t) = T(\dot{u};t) - U(u;t) \tag{2}$$

with kinetic energy

$$T(\dot{u};t) = \tfrac{1}{2} m[\dot{u}(t)]^2 \tag{3}$$

and elastic strain energy

$$U(u;t) = \tfrac{1}{2} k \left[u(t)\right]^2 \tag{4}$$

For stationary action, the first variation of (1) must be zero. Thus, the action variation $\delta A$ is

$$\delta A = -\delta \int_{t_0}^{t_1} L(u, \dot{u}; \tau) d\tau = 0 \tag{5}$$

or

$$\delta A = -\int_{t_0}^{t_1} \left[ \frac{\partial L}{\partial \dot{u}} \delta \dot{u} + \frac{\partial L}{\partial u} \delta u \right] d\tau = 0 \tag{6}$$

and finally

$$\delta A = -\int_{t_0}^{t_1} \left[ m \dot{u} \, \delta \dot{u} - k \, u \, \delta u \right] d\tau = 0 \tag{7}$$

After applying integration by parts to the first term in the integral (7), we have

$$\delta A = \int_{t_0}^{t_1} \left[ m \ddot{u} + k \, u \right] \delta u \, d\tau - \left[ m \dot{u} \, \delta u \right]_{t_0}^{t_1} = 0 \tag{8}$$

Following Hamilton (1834), in order to recover the governing equation of motion, we must invoke the condition of zero variation at the beginning and end of the time interval

$$\delta u(t_0) = 0; \quad \delta u(t_1) = 0 \tag{9}$$

Then, (8) reduces to

$$\delta A = \int_{t_0}^{t_1} \left[ m \ddot{u} + k \, u \right] \delta u \, d\tau = 0 \tag{10}$$

Finally, after allowing arbitrary variations $\delta u$ between the endpoints, we have the equation of motion for the harmonic oscillator associated with the stationarity of the action $A$ as

$$m \ddot{u} + k \, u = 0 \tag{11}$$

Of course, we also can arrive at this equation of motion by invoking the Euler-Lagrange equation

$$\frac{d}{dt} \frac{\partial L}{\partial \dot{u}} - \frac{\partial L}{\partial u} = 0 \tag{12}$$

*2.1. Characteristics*

Hamilton's principle is firstly formulated to account for particle motion, not a continuum, and is restricted to conservative systems. The main difference between Hamilton's principle and Newton's equation of motion is that Hamilton's principle is an integral equation whereas Newton's equation of motion is a differential equation. That is, it looks at the trajectory of the system as a whole, whereas the equation of motion looks only at the local trajectory.

It is more general than Newton's equation of motion that has broad applicability including electro-magnetic fields, the motion of waves, and special relativity (see Gossick, 1967; Landau and Lifshits 1975; Slawinski, 2003).

*2.2. Critical weakness*

Hamilton's principle is not compatible to other variational principles in that it does not properly use the given initial conditions. That is, Hamilton's principle assumes that the positions at the initial and final time are known, even though we have only the given initial conditions, such as displacement and velocity in strong form. This restriction is called the end-point constraint and we already checked this (9) in the harmonic oscillator example.

*2.3. Limit*

Besides this critical end-point weakness, having Hamilton's principle to embrace non-conservative system requires another scalar function, Rayleigh's dissipation function (Rayleigh, 1877).

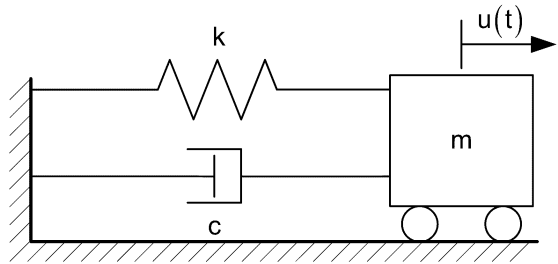

**Fig. 2. Damped oscillator**

For the damped oscillator in Fig. 2, we may define a dissipation function in the following form

$$\varphi(\dot{u};t) = \tfrac{1}{2} c \left[\dot{u}(t)\right]^2 \tag{13}$$

with the Lagrangian specified in (2)-(4).

Although the action itself no longer can be written in explicit form, the first variation of $A$ is defined as

$$\delta A = -\delta \int_{t_0}^{t_1} L(u,\dot{u};\tau)d\tau + \int_{t_0}^{t_1} \frac{\partial \varphi(\dot{u};\tau)}{\partial \dot{u}} \delta u \, d\tau = 0 \tag{14}$$

Then, from (8), (9) and (14),

$$\delta A = \int_{t_0}^{t_1} \left[ m\ddot{u} + c\dot{u} + ku \right] \delta u \, d\tau = 0 \tag{15}$$

While this approach can lead to the proper governing differential equation of motion, Rayleigh's dissipation function (13) has different dimensions (energy rate) compared to the Lagrangian (2), which deals with only energy like quantities. Also, its inconsistent first variation in (14) degenerates the completeness of variational scheme.

## 3. Previous works

### 3.1. Noether charge and Hamiltonian

To be free from the end-point constraints in the original conservative Hamilton's principle, two famous conservative quantities such as Noether charge (Noether, 1918) and Hamiltonian (Hamilton, 1835) were found. The quantity, Noether charge $\left( Q = \dfrac{\partial L}{\partial \dot{u}} \right)$, came from the assumption for the symmetry property in space of the action, and this quantity represents the generalized momentum.

On the other hand, the Hamiltonian $\left( H = \dfrac{\partial L}{\partial \dot{u}} \dot{u} - L \right)$, came from the assumption for the time translation invariance of the action. How both quantities are found is graphically represented in Fig. 3 and Fig. 4, respectively.

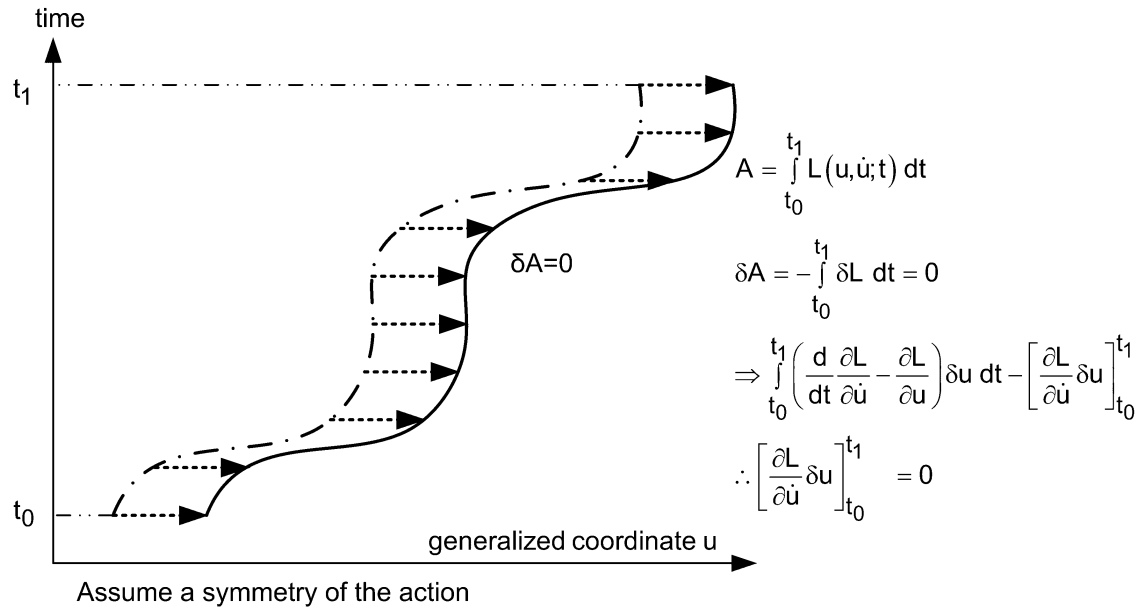

**Fig. 3.** Noether charge

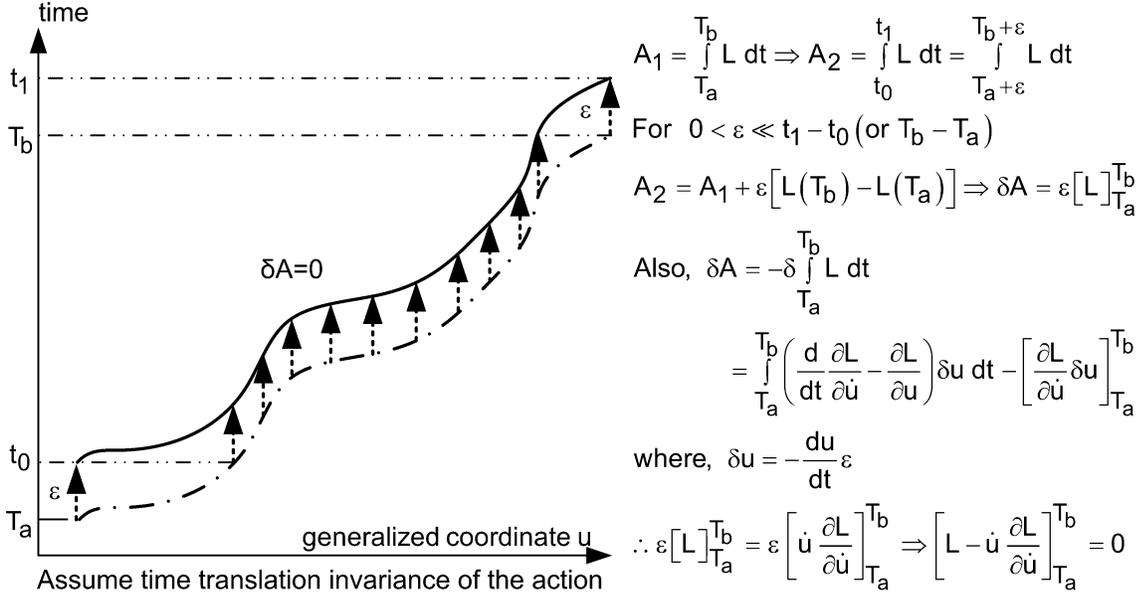

**Fig. 4. Hamiltonian**

Since initial value problems fix a certain realistic time at imaginary initial time $(t = 0)$, we may get some insight on resolving end-point constraints in Hamilton's principle from Fig. 3. When nullifying (releasing) end-point constraints in the fixed time-window in Fig. 3, the stationary statement $\delta A = 0$ considers all the trajectories where the displacement and velocity are not specified at each time boundary. In turn, if we could assign only the initial values to $\delta A$, we may have Hamilton's principle account for the initial value problem.

*3.2 Mixed Lagrangian formalism (MLF)*

Recent works by Sivaselvan and Reinhorn (2006), Lavan et al. (2008), and Sivaselvan et al. (2009) indicate a number of theoretical and computational benefits resulting from the adoption of a mixed Lagrangian formalism (MLF) for structural dynamics. Here, a framework of MLF is explained through the harmonic oscillator example.

For the harmonic oscillator in Fig. 1, MLF defines Lagrangian in mixed form:

$$L(u,\dot{u},\dot{J};t) = \frac{1}{2}m\,\dot{u}^2 + \frac{1}{2}a\,\dot{J}^2 - \dot{J}\,u \tag{16}$$

In MLF, the displacement $u(t)$ and the impulse of the spring force $J(t)$ are the primary variables, while the flexibility $a$ is used rather than the stiffness $k$. Thus, the spring force becomes:

$$F(t) = \dot{J}(t) \tag{17}$$

The flexibility and the stiffness have reciprocal relation $a = 1/k$ that the compatibility equation (Hooke's law) between the displacement and the spring force is written

$$u - a\dot{J} = 0 \tag{18}$$

With the Lagrangian (16), MLF defines the action for the time duration $[t_0, t_1]$ as

$$A = \int_{t_0}^{t_1} L(u, \dot{u}, \dot{J}; \tau) d\tau \tag{19}$$

Thus, the action variation in MLF is

$$\delta A = -\delta \int_{t_0}^{t_1} L(u, \dot{u}, \dot{J}; \tau) d\tau = 0 \tag{20}$$

or

$$\delta A = -\int_{t_0}^{t_1} \left[ \frac{\partial L}{\partial \dot{u}} \delta \dot{u} + \frac{\partial L}{\partial u} \delta u + \frac{\partial L}{\partial \dot{J}} \delta \dot{J} \right] d\tau = 0 \tag{21}$$

and finally

$$\delta A = -\int_{t_0}^{t_1} \left[ m \dot{u} \, \delta \dot{u} + a \dot{J} \, \delta \dot{J} - u \, \delta \dot{J} - \dot{J} \, \delta u \right] d\tau = 0 \tag{22}$$

After applying integration by parts to the first three terms in (22), this becomes

$$\delta A = -\left[ m \dot{u} \, \delta u \right]_{t_0}^{t_1} + \left[ \left( u - a \dot{J} \right) \delta J \right]_{t_0}^{t_1}$$
$$+ \int_{t_0}^{t_1} \left[ m \ddot{u} + \dot{J} \right] \delta u \, d\tau + \int_{t_0}^{t_1} \left[ a \ddot{J} - \dot{u} \right] \delta J \, d\tau = 0 \tag{23}$$

In (23), the first row is canceled out due to end-point constraints in MLF

$$\delta u(t_0) = 0; \quad \delta u(t_1) = 0; \quad \delta J(t_0) = 0; \quad \delta J(t_1) = 0 \tag{24}$$

Then, (23) reduces to

$$\delta A = \int_{t_0}^{t_1} \left[ \left( m \ddot{u} + \dot{J} \right) \delta u + \left( a \ddot{J} - \dot{u} \right) \delta J \right] d\tau = 0 \tag{25}$$

Finally, we have Euler-Lagrangian equations

$$m \ddot{u} + \dot{J} = 0; \quad a \ddot{J} - \dot{u} = 0 \tag{26}$$

in a coupled way of the mixed variables, each of which represents the equation of motion and rate-compatibility, valid at any time $t \, (t_0 < t < t_1)$. These are the most distinguished feature of MLF and great advantages to account for structural dynamics. That is, MLF could perceive a dynamical system as a collection of Euler-Lagrangian equations in state variables so that it provides a framework where displacement, internal forces, and other state variables can be treated uniformly.

However, it is questionable for MLF to have such end-point constraints as the last two equations in (24) since otherwise we could explicitly have the compatibility equation (18) in (23).

## 4. Extension framework of Hamilton's principle

*4.1. Sequential viewpoints for Hamilton's principle*

We may view Hamilton's principle sequentially as

1. Define Lagrangian: The system properties are defined.
2. Define action: Fix the time-window for the considered time duration.
3. Stationary of the action $\delta A = 0$: Consider all the trajectories where the displacement and velocity have arbitrary (multiple) values at initial and final time.
4. End-point constraints: Find the trajectory having the known initial and final position.

Obviously, Hamilton's principle assigns time boundary conditions to the system rather than the given initial conditions at the last step. Also, it is questionable to consider all the trajectories where the displacement and velocity have multiple values at initial and final time.

Thus, we may correctly account for the initial value problem in Hamilton's principle if Hamilton's principle has framework as

1. Define Lagrangian
2. Define action
3. **Stationary of the action** $\delta A_{NEW} = 0$: Consider the trajectories where the velocity and the displacement have unique but unspecified value at initial and final time.
4. **Assign the given initial conditions**: Find the trajectory having the known initial conditions.

In other words, we extend the action variation as $\delta A_{NEW}$, and assign the given initial values to it. This assigning process also has a sequence and this is discussed in specific examples.

## 4.2. Particle dynamics

Consider a particle having mass $m$ on a frictionless surface with $\dot{u}(t)$ shown in Fig. 5.

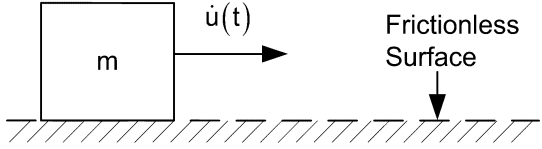

**Fig. 5. Particle motion**

The Lagrangian and the action of this system for the time duration $[0,T]$ are written

$$L(\dot{u};t) = \tfrac{1}{2} m [\dot{u}(t)]^2 \qquad (27)$$

$$A = \int_0^T L(\dot{u};\tau)d\tau \qquad (28)$$

We newly define the action variation for (28) as

$$\delta A_{NEW} = -\delta \int_0^T L(\dot{u};\tau)d\tau + \left[ m\,\hat{\dot{u}}_T\, \delta\hat{u}_T - m\,\hat{\dot{u}}_0\, \delta\hat{u}_0 \right] \qquad (29)$$

to confine our focus on the trajectories where the velocity and the displacement have one unspecified value $(\hat{\dot{u}}_0, \hat{u}_0)$ at initial and $(\hat{\dot{u}}_T, \hat{u}_T)$ at final time.

The additional closed bracket terms in (29) are nothing but the counterparts to the terms without end-point constraints in Hamilton's principle, and we will sequentially assign the known initial values $\bar{\dot{u}}_0$ and $\bar{u}_0$ to the undetermined reserved initial values $\hat{\dot{u}}_0$ and $\hat{u}_0$.

That is, (29) could be changed into

$$\begin{aligned}
\delta A_{NEW} &= -\int_0^T \left( \frac{\partial L}{\partial \dot{u}}\delta\dot{u} + \frac{\partial L}{\partial u}\delta u \right) d\tau + \left[ m\,\hat{\dot{u}}_T\,\delta\hat{u}_T - m\,\hat{\dot{u}}_0\,\delta\hat{u}_0 \right] \\
&= \int_0^T \left( \frac{d}{dt}\left(\frac{\partial L}{\partial \dot{u}}\right) - \frac{\partial L}{\partial u} \right)\delta u\, d\tau - \left[ \frac{\partial L}{\partial \dot{u}}\delta u \right]_0^T + \left[ m\,\hat{\dot{u}}_T\,\delta\hat{u}_T - m\,\hat{\dot{u}}_0\,\delta\hat{u}_0 \right]
\end{aligned} \qquad (30)$$

and we forcibly match each term of closed brackets as

$$\frac{\partial L}{\partial \dot{u}}(0) = m\,\hat{\dot{u}}_0; \quad \delta u(0) = \delta\hat{u}_0; \quad \frac{\partial L}{\partial \dot{u}}(T) = m\,\hat{\dot{u}}_T; \quad \delta u(0) = \delta\hat{u}_T \qquad (31)$$

Finally, we identify the unspecified initial values by assigning the given initial value

$$\hat{\dot{u}}_0 = \bar{\dot{u}}_0 \qquad (32)$$

and successively

$$\delta \hat{u}_0 = \delta \bar{u}_0 = 0 \quad \text{or} \quad \bar{u}_0 \text{ is given} \tag{33}$$

Since the subsequent zero-valued term (33) needs not appear explicitly in the new action variation, the new action variation definition (29) with the sequential assigning process (32)-(33) can account for the initial value problem.

This process is explained pictorially in Fig. 6 with the comparison to the original framework of Hamilton's principle. To emphasize that we only use the known initial conditions and leave the final values uniquely unknown, the circle (displacement) and the tangent line (velocity) at each end are shown in different ways.

| Framework of Hamilton's Principle | Extension Framework of Hamilton's Principle |
|---|---|
| **1. Stationary of the action ($\delta A = 0$)** | **1. Stationary of the action ($\delta A_{NEW} = 0$)** |
| 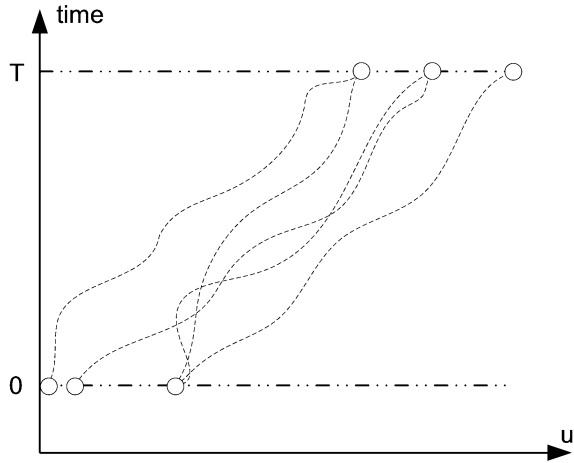 | 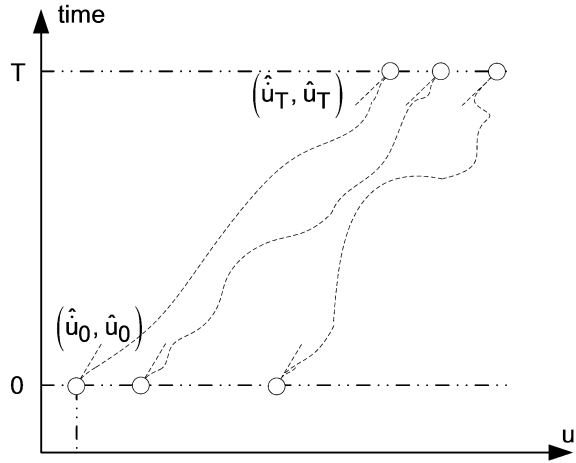 |
| Consider all the trajectories where the displacement and velocity have arbitrary (multiple) values at initial/final time | Consider the trajectories where the velocity and the displacement have unique unspecified value at initial/final time |
| **2. End-points constraint** | **2. Identify the given initial values** |
| 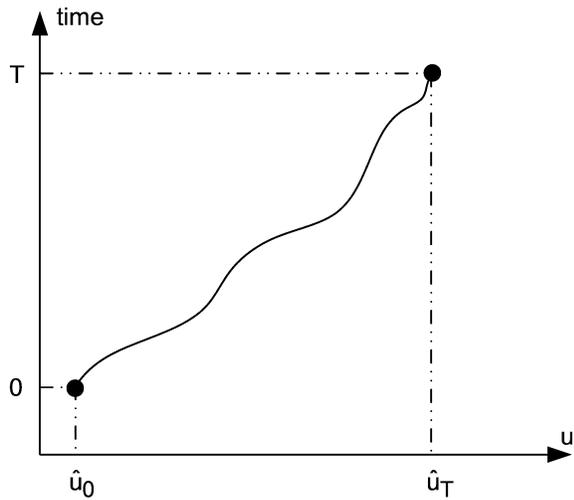 | 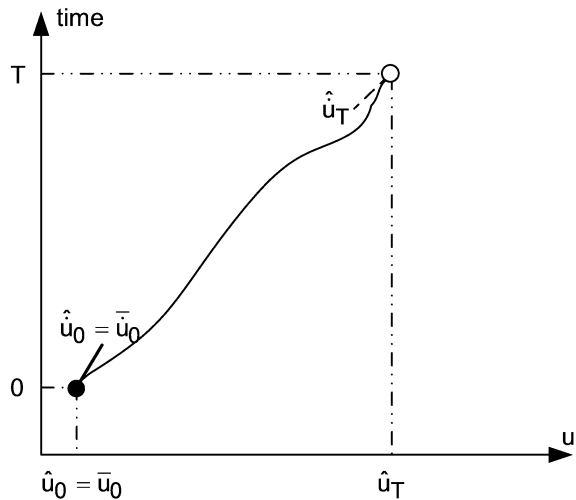 |
| Find the trajectory having the known initial and final displacement | Sequentially assign the given values $\bar{\dot{u}}_0$ and $\bar{u}_0$ to $\hat{\dot{u}}_0$ and $\hat{u}_0$, and find such trajectory. |

**Fig. 6. Graphical viewpoint for the extension framework of Hamilton's principle**

*4.3. Elastic continuum dynamics*

The extension framework is also applicable to an elastic continuum in mixed forms. Our objective is to write the new action variation that recovers all the governing equations with compatible initial and boundary conditions.

Sivaselvan and Reinhorn (2006) defines the Lagrangian density $l$ for an elastic continuum

$$l = \frac{1}{2}\rho \dot{u}_i \dot{u}_i + \frac{1}{2} A_{ijkl} \dot{J}_{ij} \dot{J}_{kl} - \dot{J}_{ij} \varepsilon_{ij} \tag{34}$$

by the generalized displacement field $u_i$ and the generalized stress field $\dot{J}_{ij}(=\sigma_{ij})$

In (34), $\rho$ is the mass density and $A_{ijkl}$ is the elastic constitutive tensor inverse to $D_{ijkl}$, the usual constitutive tensor for an anisotropic elastic medium, while

$J_{ij}(t) = \int_0^t \sigma_{ij}(\tau) d\tau$ is an impulse of stress tensor $\sigma_{ij}$ and $\varepsilon_{ij}$ is the strain tesor.

With the known body force density $\hat{f}_i$ and the known traction $\hat{t}_i$ on the portion of boundary $\Gamma_t$, we may write the applied force potential $V$

$$V = \int_\Omega \hat{f}_i u_i d\Omega + \int_{\Gamma_t} \hat{t}_i u_i d\Gamma \tag{35}$$

for an elastic continuum occupying $\Omega$ in space. Here, we assume that the boundary conditions are defined, such that $\Gamma_u \cup \Gamma_t = \Gamma$ and $\Gamma_u \cap \Gamma_t = \varnothing$.

The action for the time duration $[0,T]$ is

$$A = \int_0^T \int_\Omega l \, d\Omega \, d\tau + \int_0^T V \, d\tau \tag{36}$$

We define the new action variation for an elastic continuum in MLF framework as

$$\delta A_{NEW} = -\delta \int_0^T \int_\Omega l \, d\Omega \, d\tau - \delta \int_0^T V \, d\tau + \int_\Omega \left[ \rho \hat{\dot{u}}_i \, \delta \hat{u}_i \right]_0^T d\Omega \tag{37}$$

Equation (37) could be defined by adding all the counterparts to the terms without end-point constraints in Hamilton's principle and confining them to unique but undetermined value at initial and final time.

After performing all of the temporal integration-by-parts operations on (37), we have

$$\delta A_{NEW} = -\int_{\Omega} \left[\rho\,\dot{u}_i\,\delta u_i\right]_0^T d\Omega + \int_{\Omega}\int_0^T \rho\,\ddot{u}_i\,\delta u_i\,d\tau\,d\Omega$$

$$-\int_{\Omega}\left[A_{ijkl}\,\dot{J}_{ij}\,\delta J_{kl}\right]_0^T d\Omega + \int_{\Omega}\int_0^T A_{ijkl}\,\ddot{J}_{ij}\,\delta J_{kl}\,d\tau\,d\Omega$$

$$+\int_{\Omega}\left[\varepsilon_{ij}\,\delta J_{ij}\right]_0^T d\Omega - \int_{\Omega}\int_0^T \dot{\varepsilon}_{ij}\,\delta J_{ij}\,d\tau\,d\Omega + \int_{\Omega}\int_0^T \dot{J}_{ij}\,\delta\varepsilon_{ij}\,d\tau\,d\Omega \quad (38)$$

$$-\int_{\Omega}\int_0^T \hat{f}_i\,\delta u_i\,d\tau\,d\Omega - \int_{\Gamma_t}\int_0^T \hat{t}_i\,\delta u_i\,d\tau\,d\Gamma + \int_{\Omega}\left[\rho\,\hat{\dot{u}}_i\,\delta\hat{u}_i\right]_0^T d\Omega$$

In (38), we can also perform a spatial integration by parts on the term $\dot{J}_{ij}\delta\varepsilon_{ij}$. For this development, we make use of the symmetry of stresses $\dot{J}_{ij}$ and the Cauchy definition of surface traction, where $t_i = \dot{J}_{ij}n_j$. The reformulation for $\dot{J}_{ij}\delta\varepsilon_{ij}$ proceeds as follows

$$\int_{\Omega}\int_0^T \dot{J}_{ij}\,\delta\varepsilon_{ij}\,d\tau\,d\Omega = \int_0^T\int_{\Omega} \dot{J}_{ij}\,\delta u_{i,j}\,d\Omega\,d\tau$$

$$= \int_0^T\int_{\Omega}\left[\dot{J}_{ij}\,\delta u_i\right]_{,j} d\Omega\,d\tau - \int_0^T\int_{\Omega}\left[\dot{J}_{ij,j}\,\delta u_i\right] d\Omega\,d\tau$$

$$= \int_0^T\int_{\Gamma} \dot{J}_{ij}\,\delta u_i\,n_j\,d\Gamma\,d\tau - \int_0^T\int_{\Omega} \dot{J}_{ij,j}\,\delta u_i\,d\Omega\,d\tau \quad (39)$$

$$= \int_0^T\int_{\Gamma} t_i\,\delta u_i\,d\Gamma\,d\tau - \int_0^T\int_{\Omega} \dot{J}_{ij,j}\,\delta u_i\,d\Omega\,d\tau$$

After substituting (39) into (38), we have

$$\delta A_{NEW} = -\int_0^T\int_{\Omega}\left(\rho\,\ddot{u}_i - \dot{J}_{ij,j} - \hat{f}_i\right)\delta u_i\,d\Omega\,d\tau$$

$$+ \int_0^T\int_{\Omega}\left(A_{ijkl}\,\ddot{J}_{ij} - \dot{\varepsilon}_{kl}\right)\delta J_{kl}\,d\Omega\,d\tau$$

$$+ \int_{\Omega}\left[\underline{\left(\varepsilon_{kl} - A_{ijkl}\,\dot{J}_{ij}\right)\delta J_{kl}}\right]_0^T d\Omega \quad (40)$$

$$+ \int_0^T\int_{\Gamma} t_i\,\delta u_i\,d\Gamma\,d\tau - \int_0^T\int_{\Gamma_t} \hat{t}_i\,\delta u_i\,d\Gamma\,d\tau$$

$$+ \int_{\Omega}\left[\rho\,\hat{\dot{u}}_i\,\delta\hat{u}_i - \rho\,\dot{u}_i\,\delta u_i\right]_0^T d\Omega$$

In (40), we could explicitly have the compatibility equation (the underlined terms) for an elastic continuum without end-point constraints for the impulse. More importantly, the new action variation (37) uses all the pertinent initial/boundary conditions. That is, by expressing the displacement of an elastic continuum $u_i$ as a function of the position vector $\vec{x}$ and time $t$ as $u_i = u_i(\vec{x}, t)$, we could see that the given initial velocity condition

$$\dot{u}_i(\vec{x}, 0) = \hat{\dot{u}}_i(\vec{x}, 0) = \bar{\dot{u}}_i(\vec{x}, 0) \tag{41}$$

and successively the given initial displacement condition

$$u_i(\vec{x}, 0) = \hat{u}_i(\vec{x}, 0) = \bar{u}_i(\vec{x}, 0) \tag{42}$$

are properly used in the last line of (40).

Regarding the boundary conditions, we could also see that

$$t_i = \bar{t}_i \quad \text{on } \Gamma_t \tag{43}$$

and

$$\delta u_i(\hat{\vec{x}}, t) = 0 \quad \text{or} \quad u_i(\hat{\vec{x}}, t) = \hat{u}_i(\hat{\vec{x}}, t) \quad \text{on } \Gamma_u \tag{44}$$

are properly used in the forth line in (40).

In (44), $\hat{u}_i(\hat{\vec{x}}, t)$ is the given displacement boundary condition at specified location $\hat{\vec{x}}$.

In summary, by the new action variation, Hamilton's principle could have compatible initial conditions to the strong form. The new action variation is defined by adding the counterparts to the terms without end-point constraints in Hamilton's principle, which confines the trajectories of the dynamic system to have unknown but unique point(s) in the phase-plane at initial/final time. Regarding these additional terms as sequentially assigning the known initial values completes this extension framework for initial value problems.

By introducing Rayleigh's dissipation function to this framework, we can also account for non-conservative systems and this is shown for representative lumped parameter models, next.

## 5. Numerical implementation for lumped parameter models

In this Section, the extension framework is numerically implemented through Galerkin's method for the forced damped oscillator and elastic viscoplastic dynamics system. Since the terms such as $u$ and $\delta u$ in $\delta A_{NEW}$ could also be thought as the real displacement and the virtual displacement by virtue of extending principle of virtual work to dynamics, we do not differentiate the first variation from the virtual field from now onward. That is, $\delta A_{NEW}$ could have the weak formalism for initial value problems.

*5.1. Elasticity*

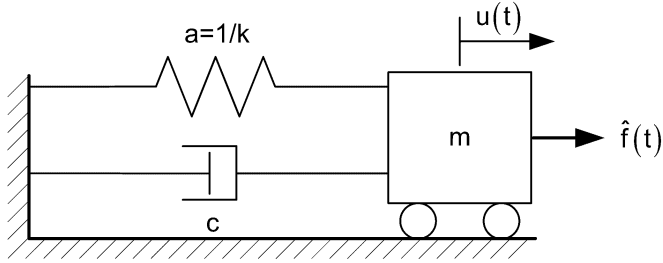

**Fig. 7. Forced damped oscillator**

With the mixed form in Lagrangian (16), the known external forcing $\hat{f}(t)$, and the Rayleigh's dissipation function (13), we define a new action variation for the forced damped oscillator displayed in Fig. 7 as

$$\delta A = -\int_0^T \left[ m\dot{u}\,\delta\dot{u} - c\dot{u}\,\delta u - \dot{J}\,\delta u + \hat{f}\,\delta u + aJ\,\delta J - u\,\delta J \right] d\tau + \left[\hat{p}\,\delta\hat{u}\right]_0^T \quad (45)$$

where,

$$\hat{p}(t) = m\,\hat{\dot{u}}(t) \quad (46)$$

A new action variation (45) is a sum of variational action integrals as $\delta A = \delta A_{1[0,t_1]} + \cdots + \delta A_{r[t_{r-1},t_r]} + \cdots + \delta A_{N[t_{N-1},t_N]}$, where $\delta A_r$ represents the action variation in the $r^{th}$ time duration $[t_{r-1}, t_r]$.

By introducing the fixed time step $h$ for each time duration, that is, $t_r = r\,h$, (45) can be written

$$\begin{aligned}\delta A &= \sum_{r=1}^N \delta A_r \\ &= \sum_{r=1}^N -\left( \int_{t_{r-1}}^{t_r} \left[ m\dot{u}\,\delta\dot{u} - c\dot{u}\,\delta u - \dot{J}\,\delta u + \hat{f}\,\delta u + aJ\,\delta J - u\,\delta J \right] d\tau - \left[\hat{p}\,\delta\hat{u}\right]_{t_{r-1}}^{t_r} \right)\end{aligned} \quad (47)$$

Equation (47) has $C^0$ time continuity for both real field and virtual field of $u$ and $J$. Thus, we could use the linear shape function to approximate each field in the action variation. For the $r^{th}$ variation of the action, the displacement field could be approximated by

$$u = \frac{1}{h}[t_r - t \quad t - t_{r-1}]\begin{Bmatrix} {}^{r-1}u \\ {}^r u \end{Bmatrix} \tag{48}$$

In (48), the new notations are used for convenience. That is, ${}^{r-1}u$ and ${}^r u$ represent the discrete value for displacement at time $t_{r-1}$ and $t_r$, respectively.

Then, $\dot{u}$, $\delta u$ and $\delta \dot{u}$ can be approximated by

$$\dot{u} = \frac{1}{h}[-1 \quad 1]\begin{Bmatrix} {}^{r-1}u \\ {}^r u \end{Bmatrix}$$

$$\delta u = \frac{1}{h}[t_r - t \quad t - t_{r-1}]\begin{Bmatrix} \delta^{r-1}u \\ \delta^r u \end{Bmatrix} \tag{49}$$

$$\delta \dot{u} = \frac{1}{h}[-1 \quad 1]\begin{Bmatrix} \delta^{r-1}u \\ \delta^r u \end{Bmatrix}$$

Similarly, $J$, $\dot{J}$, $\delta J$, and $\delta \dot{J}$ can be approximated by

$$J = \frac{1}{h}[t_r - t \quad t - t_{r-1}]\begin{Bmatrix} {}^{r-1}J \\ {}^r J \end{Bmatrix}; \quad \dot{J} = \frac{1}{h}[-1 \quad 1]\begin{Bmatrix} {}^{r-1}J \\ {}^r J \end{Bmatrix}$$

$$\delta J = \frac{1}{h}[t_r - t \quad t - t_{r-1}]\begin{Bmatrix} \delta^{r-1}J \\ \delta^r J \end{Bmatrix}; \quad \delta \dot{J} = \frac{1}{h}[-1 \quad 1]\begin{Bmatrix} \delta^{r-1}J \\ \delta^r J \end{Bmatrix} \tag{50}$$

Furthermore, let us change the forcing term $\hat{f}$ to ${}^r\hat{f}$, which represents the discrete forcing value at time $t_r$. If the forcing term $\hat{f}$ is continuous such, as a sine function or cosine function, this discretization process is not necessary, because we can put continuous function in the integral and they can be analytically evaluated. However, this discrete forcing term $t_r$ is introduced to account for general discrete forcing inputs.

Substituting (48)-(50) into (47) and doing integration yield

$$\delta A_r = -\frac{m}{h}\left({}^r u - {}^{r-1}u\right)\left(\delta^r u - \delta^{r-1}u\right) + \frac{c}{2}\left({}^r u - {}^{r-1}u\right)\left(\delta^{r-1}u + \delta^r u\right)$$
$$+ \frac{1}{2}\left({}^r J - {}^{r-1}J\right)\left(\delta^{r-1}u + \delta^r u\right) - {}^r\hat{f}\frac{h}{2}\left(\delta^{r-1}u + \delta^r u\right)$$
$$- \frac{a}{h}\left({}^r J - {}^{r-1}J\right)\left(\delta^r J - \delta^{r-1}J\right) + \frac{1}{2}\left({}^{r-1}u + {}^r u\right)\left(\delta^r J - \delta^{r-1}J\right)$$
$$+ \hat{p}(t_r)\delta^r \hat{u} - \hat{p}(t_{r-1})\delta^{r-1}\hat{u}$$
$$\tag{51}$$

Collecting terms by the virtual discrete terms $\delta^{r-1}u$, $\delta^r u$, $\delta^{r-1}J$ and $\delta^r J$ leads to

$$\delta A_r = \left[\frac{m}{h}\left({}^r u - {}^{r-1}u\right) + \frac{c}{2}\left({}^r u - {}^{r-1}u\right) + \frac{1}{2}\left({}^r J - {}^{r-1}J\right) - {}^r\hat{f}\frac{h}{2} - \hat{p}(t_{r-1})\right]\delta^{r-1}u$$

$$+ \left[-\frac{m}{h}\left({}^r u - {}^{r-1}u\right) + \frac{c}{2}\left({}^r u - {}^{r-1}u\right) + \frac{1}{2}\left({}^r J - {}^{r-1}J\right) - {}^r\hat{f}\frac{h}{2} + \hat{p}(t_r)\right]\delta^r u \quad (52)$$

$$+ \left[\frac{a}{h}\left({}^r J - {}^{r-1}J\right) - \frac{1}{2}\left({}^{r-1}u + {}^r u\right)\right]\delta^{r-1}J$$

$$+ \left[-\frac{a}{h}\left({}^r J - {}^{r-1}J\right) + \frac{1}{2}\left({}^{r-1}u + {}^r u\right)\right]\delta^r J$$

Here, we freely use the relation (42) in (52).

By letting each coefficient of the virtual discrete terms in (52) be zero, we could have three independent equations. That is, the third row equation and the fourth row equation in (52) are the same. These equations are expressed in compact matrix form as

$$\begin{bmatrix} \frac{m}{h}+\frac{c}{2} & 0 & \frac{1}{2} \\ -\frac{m}{h}+\frac{c}{2} & 1 & \frac{1}{2} \\ -\frac{1}{2} & 0 & \frac{a}{h} \end{bmatrix} \begin{Bmatrix} {}^r u \\ \hat{p}(t_r) \\ {}^r J \end{Bmatrix} = \begin{bmatrix} \frac{m}{h}+\frac{c}{2} & 1 & \frac{1}{2} \\ -\frac{m}{h}+\frac{c}{2} & 0 & \frac{1}{2} \\ \frac{1}{2} & 0 & \frac{a}{h} \end{bmatrix} \begin{Bmatrix} {}^{r-1}u \\ \hat{p}(t_{r-1}) \\ {}^{r-1}J \end{Bmatrix} + \begin{Bmatrix} \frac{h}{2}{}^r\hat{f} \\ \frac{h}{2}{}^r\hat{f} \\ 0 \end{Bmatrix} \quad (53)$$

Equation (53) is one-step time marching numerical method for the forced damped oscillator along with the given initial conditions ${}^0 u$ and $\hat{p}(t_0)$. The initial impulse ${}^0 J$ could be identified by the momentum balance equation in mixed form:

$$\hat{p}(t_0) + c\,{}^0 u + {}^0 J = {}^0\hat{I} \quad (54)$$

Here, ${}^0\hat{I}$ is the impulse of the externally applied dynamic force to the system before the initial time

$${}^0\hat{I} = \int_{-\infty}^{0} \hat{f}(\tau)\,d\tau \quad (55)$$

where $-\infty$ represents that this is the time before the time interval we consider.

If the system is initially static, the initial impulse in the spring is identified as ${}^0 J = 0$ because we have ${}^0\hat{I} = 0$ without damping and inertia effects.

## 5.2. Viscoplasticity

Consider the elastic viscoplastic dynamic system of Fig. 8.

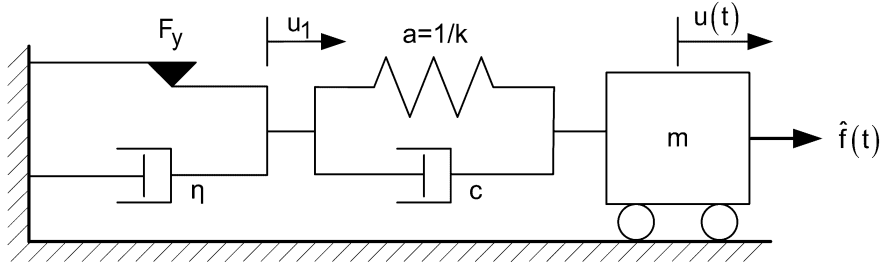

**Fig. 8. Elastic viscoplastic dynamic system**

In Fig. 8, $F_y$ and $\eta$ are the yield force and the coefficient of the regularizing viscous damper while $u_1$ is the deformation for the slider-dashpot component.

Based upon the previous results from the damped oscillator with the external forcing, we focus on the derivation of the nonlinear dissipation part. This is guaranteed because the considered model is nothing but a series of the regularization and the damped oscillator.

Previously, MLF defines the dissipation function for this model as

$$\varphi(\dot{J};t) = \frac{1}{2\eta} <|\dot{J}|-F_y>^2 \tag{56}$$

Here, $<\cdot>$ and $|\cdot|$ represent Macaulay bracket and the absolute value. The first variation of (56) with respect to $\dot{J}$ represents the plastic strain rate $\dot{u}_1$ or the rate-deformation for the slider-dashpot component

$$\dot{u}_1 = \frac{\partial \varphi(\dot{J})}{\partial \dot{J}} = \frac{1}{\eta} <|\dot{J}|-F_y> sgn(\dot{J}) \tag{57}$$

where $sgn(\cdot)$ is the signum function.

Combining (57) and the action variation terms from forced damped oscillator in (45), we have the action variation for the time duration $[t_{r-1}, t_r]$ as

$$\delta A = -\int_{t_{r-1}}^{t_r} \left[ m\ddot{u}\,\delta\dot{u} - c\dot{u}\,\delta u - \dot{J}\,\delta u + \hat{f}\,\delta u + \underline{a\dot{J}\,\delta\dot{J} - u\,\delta\dot{J} - \dot{u}_1\,\delta J} \right] d\tau + \left[\hat{p}\,\delta u\right]_{t_{r-1}}^{t_r} \tag{58}$$

It should be noted that the underlined terms in (58) does not solely represent the compatibility equation for the model in Fig. 8 because the compatibility equation is given by

$$u = a\dot{J} + u_1 \tag{59}$$

Therefore, we need to confine the deformation of the slider to be one specific undetermined value such as $\hat{u}_1(t_{r-1})$ and $\hat{u}_1(t_r)$ at initial and final time in our extension framework. Overall, we define the new action variation for the model in Fig. 8 as

$$\delta A_{NEW} = -\int_{t_{r-1}}^{t_r} \left[ m\ddot{u}\,\delta\dot{u} - c\dot{u}\,\delta u - \dot{J}\,\delta u + \hat{f}\,\delta u + a\,J\,\delta J - u\,\delta \dot{J} - \dot{u}_1\,\delta J \right] d\tau \quad (60)$$
$$+ \left[\hat{p}\,\delta u\right]_{t_{r-1}}^{t_r} - \left[\hat{u}_1\,\delta J\right]_{t_{r-1}}^{t_r}$$

Let us focus on the variational term of the dissipation function $(\delta A_\varphi)_r$ in (60), we have

$$(\delta A_\varphi)_r = \int_{t_{r-1}}^{t_r} \dot{u}_1\,\delta J\, d\tau = \int_{t_{r-1}}^{t_r} \frac{\partial \varphi(\dot{J})}{\partial \dot{J}} \delta J\, d\tau = \int_{t_{r-1}}^{t_r} \frac{1}{\eta} \underline{<|\dot{J}| - F_y > \operatorname{sgn}(\dot{J})}\,\delta J\, d\tau \quad (61)$$

For better understanding, the underlined term in (61) is visualized in Fig. 9.

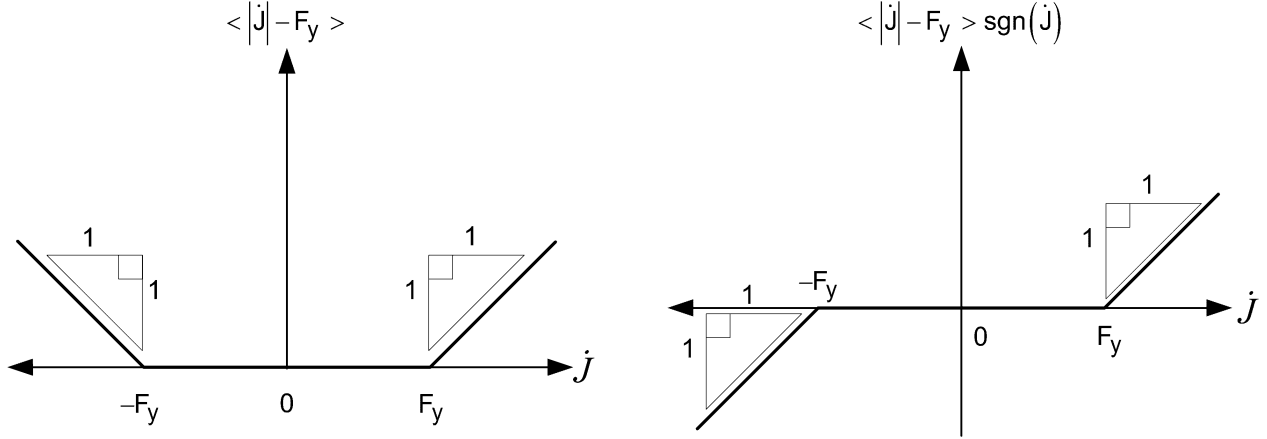

**Fig. 9. Visualize the integrand of the rate-deformation**

Approximation functions (48)-(50) are still valid since the variational form of the dissipation function does not increase or decrease these continuity requirements. Substituting (48)-(50) into the underlined function in (61) and doing the integration yield $_D(\delta A_\varphi)_r$, the discrete form of $(\delta A_\varphi)_r$

$$_D(\delta A_\varphi)_r = \begin{cases} \dfrac{h}{2\eta}\left[\dfrac{1}{h}({}^r J - {}^{r-1} J) - F_y\right](\delta^{r-1} J + \delta^r J) & \text{if } \dfrac{1}{h}({}^r J - {}^{r-1} J) > F_y \\ \dfrac{h}{2\eta}\left[\dfrac{1}{h}({}^r J - {}^{r-1} J) + F_y\right](\delta^{r-1} J + \delta^r J) & \text{if } \dfrac{1}{h}({}^r J - {}^{r-1} J) < -F_y \\ 0 & \text{otherwise (elastic case)} \end{cases} \quad (62)$$

Combining the function $_D(\delta A_\varphi)_r$ with (52) gives

$$\begin{aligned}
\delta^{r-1} u \text{ terms}: & \quad -\dfrac{h}{2}{}^r \hat{f} - \hat{p}(t_{r-1}) + \dfrac{m}{h}({}^r u - {}^{r-1} u) + \dfrac{1}{2}({}^r J - {}^{r-1} J) + \dfrac{c}{2}({}^r u - {}^{r-1} u) \\
\delta^r u \text{ terms}: & \quad -\dfrac{h}{2}{}^r \hat{f} + \hat{p}(t_r) - \dfrac{m}{h}({}^r u - {}^{r-1} u) + \dfrac{1}{2}({}^r J - {}^{r-1} J) + \dfrac{c}{2}({}^r u - {}^{r-1} u) \\
\delta^{r-1} J \text{ terms}: & \quad \dfrac{a}{h}({}^r J - {}^{r-1} J) - \dfrac{1}{2}({}^r u + {}^{r-1} u) + \hat{u}_1(t_{r-1}) + {}_D\varphi_r \\
\delta^r J \text{ terms}: & \quad -\dfrac{a}{h}({}^r J - {}^{r-1} J) + \dfrac{1}{2}({}^r u + {}^{r-1} u) - \hat{u}_1(t_r) + {}_D\varphi_r
\end{aligned} \quad (63)$$

where, $_D\varphi_r$ is

$$_D\varphi_r = \begin{cases} \dfrac{h}{2\eta}\left[\dfrac{1}{h}({}^r J - {}^{r-1} J) - F_y\right] & \text{if } \dfrac{1}{h}({}^r J - {}^{r-1} J) > F_y \\ \dfrac{h}{2\eta}\left[\dfrac{1}{h}({}^r J - {}^{r-1} J) + F_y\right] & \text{if } \dfrac{1}{h}({}^r J - {}^{r-1} J) < -F_y \\ 0 & \text{otherwise (elastic case)} \end{cases} \quad (64)$$

We have one-step time marching algorithm for viscoplasticity by making each coefficient of the discrete virtual field in (63). That is, the unknown values $\{{}^r u, {}^r J, \hat{p}(t_r), \hat{u}_1(t_r)\}$ could be found through four independent equations in (63).

The equations (63) are implicit in that they involve both the current state of the system and the subsequent one. Generally, most implicit methods are implemented into the code by using iterative methods. However, the developed method does not require any iteration, which is efficient in computation.

By introducing the notation

$$\alpha = \frac{h}{2}{}^r\hat{f}, \quad \beta = \frac{1}{2\eta}, \quad \gamma = \frac{h}{2\eta}F_y \qquad (65)$$

we can identify the time step solutions for each case as

- Elastic solutions $\left({}_D\varphi_r = 0\right)$

$$\begin{bmatrix} \frac{m}{h}+\frac{c}{2} & \frac{1}{2} & 0 & 0 \\ -\frac{m}{h}+\frac{c}{2} & \frac{1}{2} & 1 & 0 \\ -\frac{1}{2} & \frac{a}{h} & 0 & 0 \\ \frac{1}{2} & -\frac{a}{h} & 0 & -1 \end{bmatrix} \begin{Bmatrix} {}^ru \\ {}^rJ \\ \hat{p}(t_r) \\ \hat{u}_1(t_r) \end{Bmatrix} = \begin{bmatrix} \frac{m}{h}+\frac{c}{2} & \frac{1}{2} & 1 & 0 \\ -\frac{m}{h}+\frac{c}{2} & \frac{1}{2} & 0 & 0 \\ \frac{1}{2} & \frac{a}{h} & 0 & -1 \\ -\frac{1}{2} & -\frac{a}{h} & 0 & 0 \end{bmatrix} \begin{Bmatrix} {}^{r-1}u \\ {}^{r-1}J \\ \hat{p}(t_{r-1}) \\ \hat{u}_1(t_{r-1}) \end{Bmatrix} + \begin{Bmatrix} \alpha \\ \alpha \\ 0 \\ 0 \end{Bmatrix} \qquad (66)$$

- Plasticity solutions $\left({}_D\varphi_r = \frac{h}{2\eta}\left[\frac{1}{h}\left({}^rJ - {}^{r-1}J\right) - F_y\right]\right)$

$$\begin{bmatrix} \frac{m}{h}+\frac{c}{2} & \frac{1}{2} & 0 & 0 \\ -\frac{m}{h}+\frac{c}{2} & \frac{1}{2} & 1 & 0 \\ -\frac{1}{2} & \frac{a}{h}+\beta & 0 & 0 \\ \frac{1}{2} & -\frac{a}{h}+\beta & 0 & -1 \end{bmatrix} \begin{Bmatrix} {}^ru \\ {}^rJ \\ \hat{p}(t_r) \\ \hat{u}_1(t_r) \end{Bmatrix} = \begin{bmatrix} \frac{m}{h}+\frac{c}{2} & \frac{1}{2} & 1 & 0 \\ -\frac{m}{h}+\frac{c}{2} & \frac{1}{2} & 0 & 0 \\ \frac{1}{2} & \frac{a}{h}+\beta & 0 & -1 \\ -\frac{1}{2} & -\frac{a}{h}+\beta & 0 & 0 \end{bmatrix} \begin{Bmatrix} {}^{r-1}u \\ {}^{r-1}J \\ \hat{p}(t_{r-1}) \\ \hat{u}_1(t_{r-1}) \end{Bmatrix} + \begin{Bmatrix} \alpha \\ \alpha \\ \gamma \\ \gamma \end{Bmatrix} \qquad (67)$$

- Plasticity solutions $\left({}_D\varphi_r = \frac{h}{2\eta}\left[\frac{1}{h}\left({}^rJ - {}^{r-1}J\right) + F_y\right]\right)$

$$\begin{bmatrix} \frac{m}{h}+\frac{c}{2} & \frac{1}{2} & 0 & 0 \\ -\frac{m}{h}+\frac{c}{2} & \frac{1}{2} & 1 & 0 \\ -\frac{1}{2} & \frac{a}{h}+\beta & 0 & 0 \\ \frac{1}{2} & -\frac{a}{h}+\beta & 0 & -1 \end{bmatrix} \begin{Bmatrix} {}^ru \\ {}^rJ \\ \hat{p}(t_r) \\ \hat{u}_1(t_r) \end{Bmatrix} = \begin{bmatrix} \frac{m}{h}+\frac{c}{2} & \frac{1}{2} & 1 & 0 \\ -\frac{m}{h}+\frac{c}{2} & \frac{1}{2} & 0 & 0 \\ \frac{1}{2} & \frac{a}{h}+\beta & 0 & -1 \\ -\frac{1}{2} & -\frac{a}{h}+\beta & 0 & 0 \end{bmatrix} \begin{Bmatrix} {}^{r-1}u \\ {}^{r-1}J \\ \hat{p}(t_{r-1}) \\ \hat{u}_1(t_{r-1}) \end{Bmatrix} + \begin{Bmatrix} \alpha \\ \alpha \\ -\gamma \\ -\gamma \end{Bmatrix} \qquad (68)$$

A non-iterative numerical algorithm could be obtained through the fact that the plasticity direction is fixed from the elastic-assumed solution $^r J^E$. That is, $^r J^E$ gives the direction of plasticity, if it occurs. The algorithm is given by

1. Find the elastic assumed solution $^r J^E$ from (66)
2. Check yield criteria
    i) If $^{r-1}J - h F_y \leq {^r J^E} \leq {^{r-1}J} + h F_y$: Elastic assumed solutions are solutions
    
    ii) If $^r J^E > {^{r-1}J} + h F_y$: Solutions from solving (67)
    
    iii) If $^r J^E < {^{r-1}J} - h F_y$: Solutions from solving (68)

Note that the new method gives the additional information, such as the residual displacement $\hat{u}_1(t_N)$ and the momentum $\hat{p}(t_N)$ at the final time $t_N$. This is necessary and sufficient information to account for further dynamics of the system.

*5.3. Numerical properties of the new method in elasticity*

To check numerical properties of the new method, consider the damped oscillator example developed in Section 5.1. Equation (53) tells that it can account for superposition in elasticity. This is so because matrix multiplication is distributive:

Let $B$ be the left side matrix and $D$ be the right side matrix in (53). Furthermore, let $y$ be the unknown vector solution and $x$ be the known vector solution along with that $z$ is the right hand side known vector. Then, (53) is simplified to

$$B y = D x + z \tag{69}$$

Assume that the vector $x_1$, $y_1$ and $z_1$ satisfy the equation:

$$B y_1 = D x_1 + z_1 \tag{70}$$

Also, assume that the vector $x_2$, $y_2$ and $z_2$ satisfy the equation:

$$B y_2 = D x_2 + z_2 \tag{71}$$

Then, the solution $y_3$ for the equation

$$B y_3 = D(x_1 + x_2) + (z_1 + z_2) \tag{72}$$

is

$$y_3 = (x_1 + x_2) \tag{73}$$

The left side matrix $B$ always has its inverse. That is, we have the inverse matrix as

$$\begin{bmatrix} \dfrac{m}{h}+\dfrac{c}{2} & 0 & \dfrac{1}{2} \\ -\dfrac{m}{h}+\dfrac{c}{2} & 1 & \dfrac{1}{2} \\ -\dfrac{1}{2} & 0 & \dfrac{a}{h} \end{bmatrix}^{-1} = \begin{bmatrix} \dfrac{4\,a\,h}{X} & 0 & -\dfrac{2\,h^2}{X} \\ -\dfrac{\left(h^2+2\,a\,c\,h-4\,a\,m\right)}{X} & 1 & -\dfrac{4\,m\,h}{X} \\ \dfrac{2\,h^2}{X} & 0 & \dfrac{2(2\,m+c\,h)\,h}{X} \end{bmatrix} \qquad (74)$$

where,

$$X = h^2 + 2\,a\,c\,h + 4\,a\,m \qquad (75)$$

With the physical quantity for $a$, $c$, $m$, and $h$ (that is, $a>0$, $c>0$, $m>0$ and $h>0$), $X$ is always greater than zero so that $B^{-1}$ always exists.

Also, $B^{-1}D$ is simplified to

$$\dfrac{1}{X}\begin{bmatrix} 4\,m\,a+2\,a\,c\,h-h^2 & 4\,a\,h & 0 \\ -4\,m\,h & Y & 0 \\ 2(2\,m+c\,h)\,h & 2\,h^2 & X \end{bmatrix} \qquad (76)$$

where,

$$Y = 4\,m\,a - 2\,a\,c\,h - h^2 \qquad (77)$$

All the eigenvalues of (76) have the magnitude less than or equal to one for the under-damped system (that is, $a\,c^2 < 4\,m$) that the new method is unconditionally stable.

Furthermore, the sum of the first row and the second row of (53) yields the equation of motion over the time duration $[t_{r-1},t_r]$ as

$$\dfrac{\hat{p}(t_r)-\hat{p}(t_{r-1})}{h} + \dfrac{c}{h}\left(^r u - ^{r-1} u\right) + \dfrac{\left(^r J - ^{r-1} J\right)}{h} = {^r}\hat{f} \qquad (78)$$

and the difference of the first row and the second row of (53) yields the linear velocity relation

$$m\left(^r u - ^{r-1} u\right) = \dfrac{h}{2}\left[\hat{p}(t_r) + \hat{p}(t_{r-1})\right] \qquad (79)$$

Since the new method uses the equilibrium (78) and the constant acceleration (79) over each time interval, the numerical properties of the new method are same as Newmark's constant acceleration method when accounting for the damped oscillator (see Chopra, 1995; Newmark, 1959). Thus, the new method cannot help but have symplecticity and numerical dispersion property same as Newmark's constant acceleration method when accounting for the conservative harmonic oscillator.

*5.4. Numerical properties of the new method in visoplasticity*

Similarly, we can check the numerical properties of the new method for lumped viscoplasticity model. Equations (66)-(68) can be simply expressed as

$$B_e y_e = D_e x + z_e \tag{80}$$

$$B_p y_p = V_p x + z_{p_1} \tag{81}$$

$$B_p y_p = V_p x + z_{p_2} \tag{82}$$

where, $B_e$, $D_e$, $z_e$, $B_p$, $D_p$, $z_{p_1}$, and $z_{p_2}$ are

$$B_e = \begin{bmatrix} \frac{m}{h}+\frac{c}{2} & \frac{1}{2} & 0 & 0 \\ -\frac{m}{h}+\frac{c}{2} & \frac{1}{2} & 1 & 0 \\ -\frac{1}{2} & \frac{a}{h} & 0 & 0 \\ \frac{1}{2} & -\frac{a}{h} & 0 & -1 \end{bmatrix}, \quad D_e = \begin{bmatrix} \frac{m}{h}+\frac{c}{2} & \frac{1}{2} & 1 & 0 \\ -\frac{m}{h}+\frac{c}{2} & \frac{1}{2} & 0 & 0 \\ \frac{1}{2} & \frac{a}{h} & 0 & -1 \\ -\frac{1}{2} & -\frac{a}{h} & 0 & 0 \end{bmatrix}, \quad z_e = \begin{Bmatrix} \alpha \\ \alpha \\ 0 \\ 0 \end{Bmatrix} \tag{83}$$

$$B_p = \begin{bmatrix} \frac{m}{h}+\frac{c}{2} & \frac{1}{2} & 0 & 0 \\ -\frac{m}{h}+\frac{c}{2} & \frac{1}{2} & 1 & 0 \\ -\frac{1}{2} & \frac{a}{h}+\beta & 0 & 0 \\ \frac{1}{2} & -\frac{a}{h}+\beta & 0 & -1 \end{bmatrix}, \quad D_p = \begin{bmatrix} \frac{m}{h}+\frac{c}{2} & \frac{1}{2} & 1 & 0 \\ -\frac{m}{h}+\frac{c}{2} & \frac{1}{2} & 0 & 0 \\ \frac{1}{2} & \frac{a}{h}+\beta & 0 & -1 \\ -\frac{1}{2} & -\frac{a}{h}+\beta & 0 & 0 \end{bmatrix}, \quad z_{p_1} = \begin{Bmatrix} \alpha \\ \alpha \\ \gamma \\ \gamma \end{Bmatrix}, \quad z_{p_2} = \begin{Bmatrix} \alpha \\ \alpha \\ -\gamma \\ -\gamma \end{Bmatrix} \tag{84}$$

Again, the left side matrices $B_e$, $B_p$ always have inverse for the real parameters $a$, $c$, $m$, $h$, and $\eta$ since we have

$$[B_e]^{-1} = \begin{bmatrix} \dfrac{4\,a\,h}{X} & 0 & -\dfrac{2\,h^2}{X} & 0 \\ \dfrac{2\,h^2}{X} & 0 & \dfrac{2\,h(2\,m+c\,h)}{X} & 0 \\ \dfrac{Y}{X} & 1 & -\dfrac{4\,m\,h}{X} & 0 \\ 0 & 0 & -1 & -1 \end{bmatrix} \tag{85}$$

$$[B_p]^{-1} = \begin{bmatrix} \dfrac{2\,h(h+2\,a\,\eta)}{\eta X + 2\,h\,m + c\,h^2} & 0 & -\dfrac{2\,h^2\,\eta}{\eta X + 2\,h\,m + c\,h^2} & 0 \\ \dfrac{2\,h^2\,\eta}{\eta X + 2\,h\,m + c\,h^2} & 0 & \dfrac{2\,h\,\eta(2\,m+c\,h)}{\eta X + 2\,h\,m + c\,h^2} & 0 \\ \dfrac{\eta Y - c\,h^2 + 2\,h\,m}{\eta X + 2\,h\,m + c\,h^2} & 1 & -\dfrac{4\,h\,m\,\eta}{\eta X + 2\,h\,m + c\,h^2} & 0 \\ \dfrac{2\,h^2}{\eta X + 2\,h\,m + c\,h^2} & 0 & \dfrac{-\eta X + 2\,h\,m + c\,h^2}{\eta X + 2\,h\,m + c\,h^2} & -1 \end{bmatrix} \tag{86}$$

All the eigenvalues for the matrix $[B_e]^{-1}[D_e]$ have the magnitude less than or equal to one for the under-damped system ($a\,c^2 < 4\,m$). Also, all the eigenvalues for the matrix $[B_p]^{-1}[D_p]$ have the magnitude less than or equal to one for

$$0 < c < \frac{m}{a\,\eta} + 2\sqrt{\frac{m}{a}} \tag{87}$$

Thus, as long as we consider the under-damped for elasticity ($a\,c^2 < 4\,m$ or $0 < c < 2\sqrt{\dfrac{m}{a}}$), we can always guarantee that the solutions (67)-(68) are always unconditionally stable.

## 6. Numerical simulation for a viscoplasic model

The new method for the model in Fig. 8 is tested for the initially static system with zero values for both initial displacement and velocity with the comparison to the results from OpenSees by McKenna et al. (2010).

*6.1. Simulation cases*

Two loading cases are considered for numerical experiments. First one is a sine force with resonant frequency of the elastic region in the system (called resonant loading from now on), and the other is 1940 El-Centro loading.

The fixed system properties are: $m = 1 (kip \sec^2 / in)$; $k = 225 (kip / in)$; and $F_y = 0.27 (kip / in)$.

Also, the parameters and loading characteristics for each analysis are summarized below:

**Table 1. Numerical simulation properties for each loading case**

|  | Resonant $[\sin(15t)]$ | El-Centro |
|---|---|---|
| Parameter(s) changes | Time step | $\eta$ and $c$ |
|  | 1) $h = 0.02$ (sec) | 1) $\eta = c = 1.5$ |
|  | 2) $h = 0.01$ (sec) | 2) $\eta = c = 0.9$ |
|  | 3) $h = 0.005$ (sec) | 3) $\eta = c = 0.3$ |
| Parameter(s) fixed | $\eta = c = 1.5$ | $h = 0.02$ (sec) |
| Scale factor | 0.2 | 2 |
| Loading duration (sec) | 30 | 31.16 |
| Analysis time (sec) | 40 | 40 |

In OpenSees software (ver.2.2.1), a zeroLength element with uniaxial hardening material with zero hardening ratio is used to model Fig. 8 where Newmark's constant acceleration method with Newton-Raphson's iteration (positive force convergence: the 2-norm of the displacement increment in the linear system of equations is less than the tolerance $10^{-7}$ with maximum number of iterations, 10) is used.

*6.2. Simulation results*

Displacement history and hysteresis for each loading analysis are presented, where all the results from the new developed method are illustrated by the dotted lines while the solid lines are from Newmark's constant acceleration with Newton-Raphson's iteration analysis.

Since the new method dealt primarily with the displacement and the impulse, the following relations are used for hysteresis results:

$$^r u_a = \frac{1}{2}\left(^{r-1}u + {}^r u\right) \qquad (88)$$

$$^rF = \frac{1}{h}\left(^rJ - {}^{r-1}J\right) \tag{89}$$

Here, $^ru_a$ and $^rF$ are the average displacement and the representative internal force for the time duration $[t_{r-1}, t_r]$.

From resonant loading analysis results in Fig. 10-Fig. 11, we can see that Newmark constant acceleration method is relatively sensitive to the time-step. That is, in Fig. 11, the solid lines are thicker than the dotted lines in the most coarse time-step analysis (This can also be checked in Fig. 10).

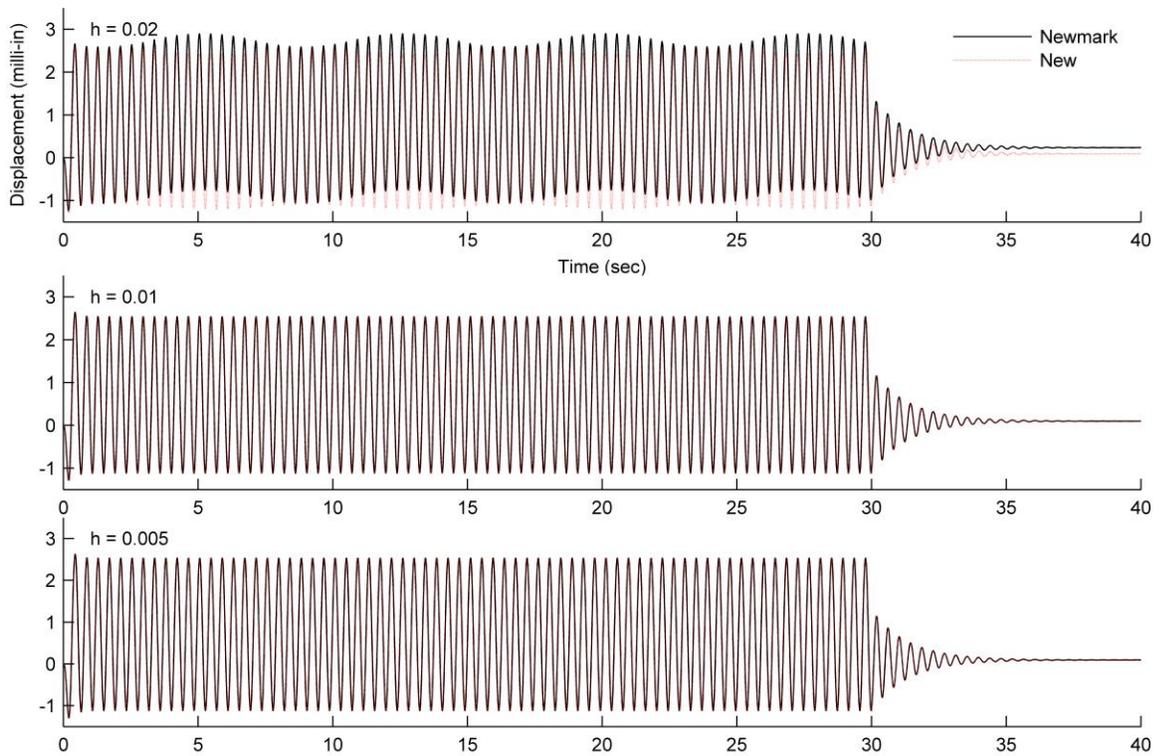

**Fig. 10. Displacement history results for resonant loading analysis**

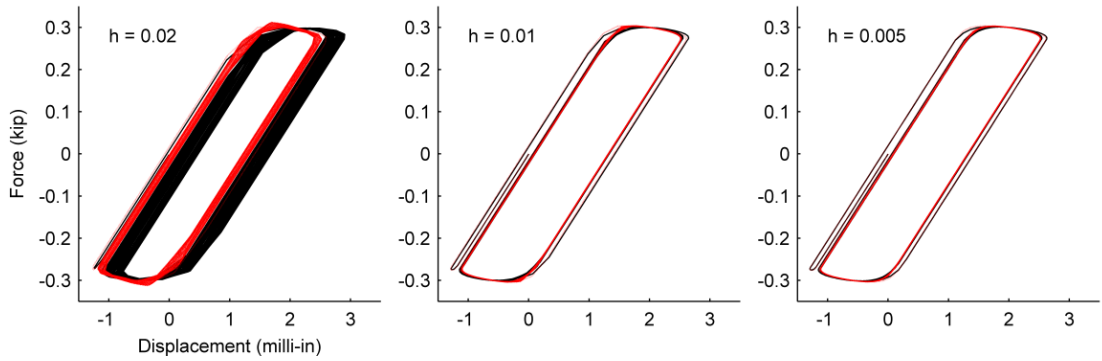

**Fig. 11. Hysteresis results for resonant loading analysis**

For the El-Centro loading analysis results in Fig. 12-Fig. 13, we could hardly see the difference between both methods. Despite the slight difference between both methods near the plastic region in Fig. 13, overall results are well matched to each other.

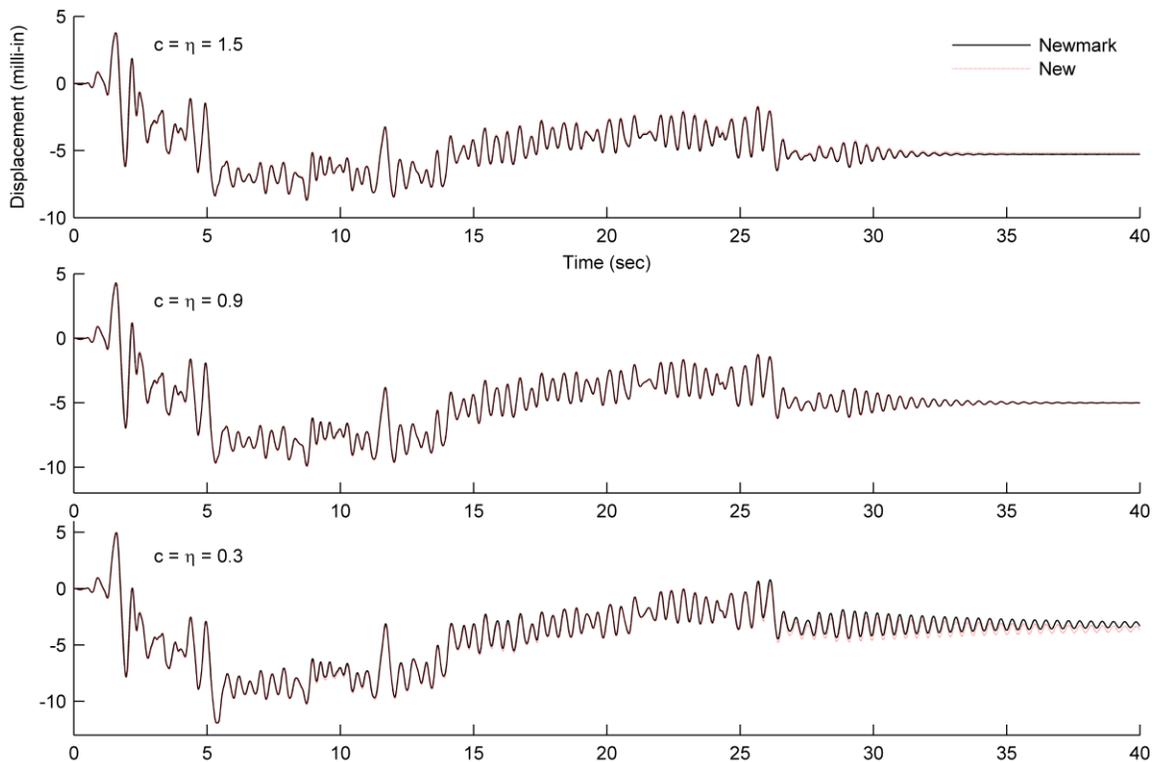

**Fig. 12. Displacement history for El-Centro loading analysis**

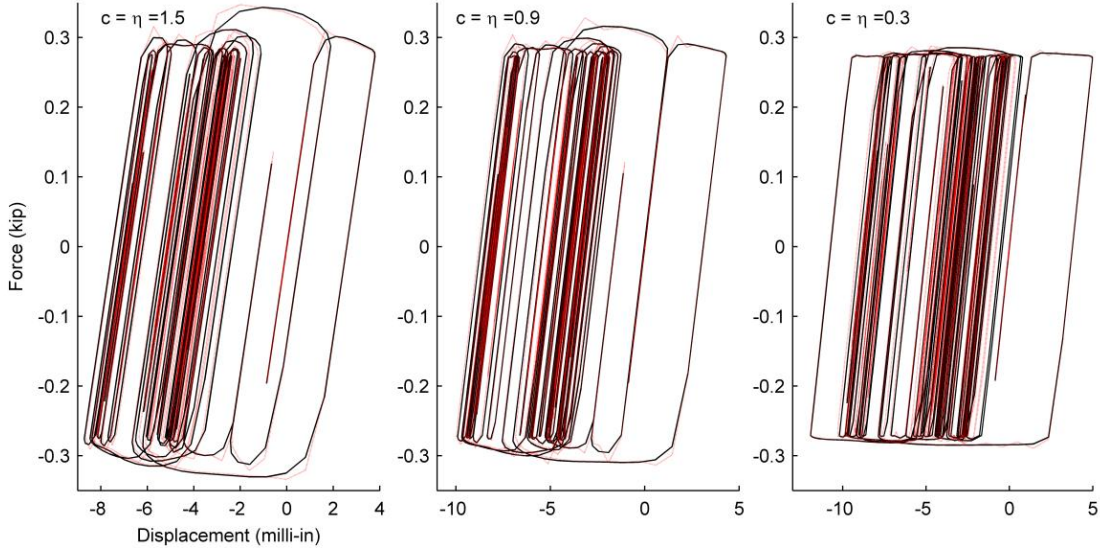

**Fig. 13. Hysteresis results for El-Centro loading analysis**

Notice that we have the additional information $\{\hat{u}_1(t_N), \hat{p}(t_N)\}$ from the new method when solving (66)-(68) at the final step $t_N$, which are sufficient and enough information for analyzing the further dynamics of the system. In our numerical examples, we have:

**Table 2. Additional results from the new method**

| Analysis case | Parameter | $\hat{u}_1(t_N)$ (milli-in) | $\hat{p}(t_N)$ (milli-in/sec) |
| --- | --- | --- | --- |
| Resonant loading analysis | $h=0.02$ | 0.0924 | -0.0120 |
|  | $h=0.01$ | 0.0965 | -0.0085 |
|  | $h=0.005$ | 0.0909 | -0.0072 |
| El-Centro loading analysis | $\xi=0.05$ | -5.1638 | -0.0054 |
|  | $\xi=0.03$ | -5.0293 | -0.1417 |
|  | $\xi=0.01$ | -3.5814 | -3.1699 |

## 7. Conclusions

In this paper, it is shown that how the critical weakness in the original Hamilton's principle could be removed by the new formulation. That is, by newly defining the action variation and sequentially assigning the initial values, non-compatible initial condition issues in Hamilton's principle could be resolved. It is not a complete variational principle since it cannot have a functional action to derive the action variation explicitly and it also requires Rayleigh's dissipation function to account for non-conservative systems. Despite such incompleteness, the framework to define the new action variation is simple, and also applicable to continuum dynamics.

The new formulation is numerically implemented for a lumped parameter viscoplastic model through extending Galerkin's finite element method into time domain. The newly developed numerical method is non-iterative, and could give whole information for the further dynamics of the system. Through some numerical simulation, it is found that the new method is compatible with Newmark constant acceleration method with Newton-Raphson's iteration.

Also, it is analytically shown that how the newly developed method has same numerical properties as Newmark's constant acceleration method when accounting for single degree of freedom oscillators.

## Acknowledgements

This research was done under the guidance by my adviser Professor Dargush (Mechanical and Aerospace Engineering, SUNY at Buffalo) during my PhD studies. The author deeply appreciates for his advice and help.